\documentclass[conference]{IEEEtran}

\usepackage{cite}
\usepackage{graphicx}
\usepackage{amsmath, amssymb}
\usepackage{algorithm}
\usepackage{algpseudocode}
\usepackage{booktabs}
\usepackage{multirow}
\usepackage{tabularx}

\begin{document}

\title{Zero-Trust Mobility-Aware Authentication Framework for Secure Vehicular Fog Computing Networks}

\author{
    \IEEEauthorblockN{Taimoor Ahmad}
    \IEEEauthorblockA{dept. of Computer Science \\
    The Superior Univeristy Lahore\\
Lahore, Pakistan \\
    Taimoor.ahmad1@superior.edu.pk}

}

\maketitle

\begin{abstract}
Vehicular Fog Computing (VFC) is a promising paradigm to meet the low-latency and high-bandwidth demands of Intelligent Transportation Systems (ITS). However, dynamic vehicle mobility and diverse trust boundaries introduce critical security challenges. This paper presents a novel Zero-Trust Mobility-Aware Authentication Framework (ZTMAF) for secure communication in VFC networks. The framework employs context-aware authentication with lightweight cryptographic primitives, a decentralized trust evaluation system, and fog node-assisted session validation to combat spoofing, replay, and impersonation attacks. Simulation results on NS-3 and SUMO demonstrate improved authentication latency, reduced computational overhead, and better scalability compared to traditional PKI and blockchain-based models. Our findings suggest that ZTMAF is effective for secure, real-time V2X interactions under adversarial and mobility-variant scenarios.

\end{abstract}

\section{Introduction}
The convergence of vehicular networks and fog computing has created a new frontier for low-latency, real-time intelligent transportation systems (ITS). Vehicular Fog Computing (VFC) allows offloading tasks from vehicles to nearby fog nodes or roadside units (RSUs), enabling applications such as collision avoidance, traffic congestion management, and autonomous driving assistance~\cite{hou2016vehicular,z3,z72}. However, the mobility of vehicles, heterogeneity of devices, and lack of persistent connections introduce critical security and privacy vulnerabilities.

One of the primary concerns in VFC is establishing trust and secure communication in a decentralized and dynamic environment. Traditional Public Key Infrastructure (PKI) systems~\cite{raya2007securing,z33} or blockchain-based methods~\cite{yang2021blockchain,z5,z74} often suffer from scalability issues, high computational overhead, and delayed response time due to consensus operations. Moreover, these methods assume static trust zones, which do not reflect the reality of vehicles moving across heterogeneous fog domains.

To address these challenges, this paper introduces a Zero-Trust Mobility-Aware Authentication Framework (ZTMAF) tailored for VFC networks. Our framework integrates context-aware session authentication with a lightweight trust evaluation engine at the fog layer. Inspired by the zero-trust principle~\cite{rose2020zero,z33333,z71}, ZTMAF eliminates implicit trust by continuously verifying entities based on behavior, context, and cryptographic credentials.

The specific research problem addressed in this paper is: How can we achieve low-latency, scalable, and secure authentication for vehicles in a dynamic fog-based network without relying on static trust assumptions?

Solving this problem is crucial for real-world deployment of ITS applications that require rapid authentication and resilience against adversarial threats. For instance, emergency braking alerts, lane-change warnings, or coordinated platooning rely on secure and timely message exchanges\cite{z55,z73}.

Our solution, ZTMAF, builds on a three-tier model comprising vehicles, fog nodes, and a decentralized trust ledger. It supports mobility-aware authentication using session keys, rolling trust scores, and context verification (speed, location, behavior). Unlike prior works, our approach does not assume stable infrastructure or global synchronization.

\textbf{Key Contributions:}
\begin{itemize}
  \item We propose ZTMAF, a novel zero-trust framework for secure, context-aware authentication in VFC networks.
  \item We design a trust evaluation algorithm based on recent behavioral data and contextual metrics such as mobility patterns.
  \item We implement a lightweight authentication protocol that minimizes computational and communication overhead, suitable for resource-constrained vehicles.
  \item We conduct extensive simulations using NS-3 and SUMO, comparing our approach against PKI and blockchain models, demonstrating superior performance in latency, scalability, and attack resilience.
\end{itemize}

The rest of the paper is organized as follows: Section II reviews related literature. Section III describes the proposed system model and authentication protocol. Section IV details the simulation setup and presents performance results. Section V concludes the paper and outlines future directions.

\section{Related Work}
Recent research efforts have focused on addressing authentication, trust, and security challenges in vehicular fog and edge computing systems. This section reviews ten significant works and highlights their key contributions and limitations.

\textbf{Li et al.~\cite{li2021secure}} proposed a blockchain-based decentralized trust management system for vehicular networks. Their approach utilizes smart contracts and distributed ledgers to ensure integrity and traceability. However, the blockchain consensus introduces latency and energy overhead unsuitable for real-time applications.

\textbf{Liu et al.~\cite{liu2020efficient}} developed an efficient and privacy-preserving authentication protocol based on elliptic curve cryptography (ECC). While their method reduces computation, it does not consider dynamic vehicle mobility or trust variations over time.

\textbf{Zhang et al.~\cite{zhang2019blockchain}} implemented a lightweight identity verification system using blockchain and fog nodes. Their protocol provides auditability but lacks adaptability to rapidly changing topologies in vehicular environments.

\textbf{Huang et al.~\cite{huang2021secure}} designed a hybrid key management scheme for V2X using identity-based encryption and fog-layer delegation. It enhances scalability, but fog nodes can become bottlenecks or single points of failure.

\textbf{Kang et al.~\cite{kang2019blockchain}} introduced reputation-aware trust models embedded in blockchain for vehicular security. They address insider threats but suffer from high storage and bandwidth costs.

\textbf{Shao et al.~\cite{shao2020efficient}} proposed a fast mutual authentication method for vehicular edge computing using one-time session keys. Though efficient, it does not support continuous trust adaptation or contextual decision-making.

\textbf{Alzahrani et al.~\cite{alzahrani2020secure}} presented a trust-based intrusion detection system for fog-assisted vehicular networks. They consider malicious behavior patterns but lack real-time enforcement mechanisms.

\textbf{Chen et al.~\cite{chen2022decentralized}} built a decentralized zero-trust security architecture tailored for fog-based Internet of Vehicles. Their system supports fine-grained access control but relies on pre-defined trust anchors.

\textbf{Wang et al.~\cite{wang2020collaborative}} introduced a collaborative authentication protocol using fog-to-fog mutual attestation. The model reduces central dependencies but incurs synchronization overhead.

\textbf{Feng et al.~\cite{feng2021mobility}} proposed a mobility-aware trust evaluation model for edge computing in vehicular networks. It dynamically adjusts trust scores based on mobility and encounter frequency, but lacks full integration with cryptographic authentication.

Most existing works either rely on static trust models or involve heavyweight consensus mechanisms that limit scalability and real-time performance. Few systems adaptively integrate contextual information like mobility or behavior for authentication. In contrast, our ZTMAF framework introduces a lightweight, context-aware, zero-trust authentication mechanism that dynamically evaluates trust and ensures secure session validation across heterogeneous vehicular fog domains.

\section{System Model}
To model our Zero-Trust Mobility-Aware Authentication Framework (ZTMAF) for secure vehicular fog computing, we construct a dynamic, decentralized network using formal graph theory and contextual trust metrics.

\subsection*{Network Abstraction and Entities}
We define the vehicular fog computing environment as an undirected graph $\mathcal{G} = (\mathcal{N}, \mathcal{L})$, where:
\begin{itemize}
  \item $\mathcal{N} = \{u_1, u_2, ..., u_M\} \cup \{f_1, f_2, ..., f_K\}$ is the set of all nodes, where $u_i$ are vehicles and $f_j$ are fog nodes,
  \item $\mathcal{L}$ is the set of authenticated communication links between nodes,
  \item $\mathcal{F} \subset \mathcal{N}$ is the subset of fog nodes, and $\mathcal{U} = \mathcal{N} \setminus \mathcal{F}$ is the set of vehicles.
\end{itemize}

Each vehicle $u_i$ interacts with nearby fog nodes $f_j$ to authenticate itself and establish a secure session. Mobility, trustworthiness, and contextual behavior are dynamically modeled to influence authentication decisions.

\subsection*{Context and Trust Evaluation}
Let $\mathbf{c}_i(t) = [s_i(t), l_i(t), b_i(t)]$ denote the context vector for vehicle $u_i$ at time $t$, capturing:
\begin{itemize}
  \item $s_i(t)$: vehicle speed,
  \item $l_i(t)$: current geolocation,
  \item $b_i(t)$: recent behavior score from local observation.
\end{itemize}

The trust score $\mathcal{T}_i(t)$ is updated using a context-sensitive exponential filter:
\begin{equation}
\mathcal{T}_i(t+1) = \alpha \cdot \mathcal{T}_i(t) + (1 - \alpha) \cdot \psi(\mathbf{c}_i(t))
\end{equation}
where $0 < \alpha < 1$ is a forgetting factor, and $\psi(\cdot)$ maps context to risk-weighted trust.

\subsection*{Authentication Request Generation}
Each vehicle generates a session request $\mathcal{R}_i$ as:
\begin{align}
\mathcal{R}_i &= H(ID_i || \mathbf{c}_i(t) || \mathcal{T}_i(t)) \\
\sigma_i &= Sign_{\mathcal{K}_{priv}^i}(\mathcal{R}_i)
\end{align}
Here, $H(\cdot)$ is a collision-resistant hash function and $\sigma_i$ is the vehicle's digital signature over the request using its private key.

\subsection*{Fog Node Verification and Session Setup}
Upon receiving $\mathcal{R}_i$, fog node $f_j$ performs:
\begin{align}
Verify(\sigma_i, ID_i, \mathcal{R}_i) &\rightarrow \text{accept/reject} \\
P_{accept} &= \mathbb{P}(\mathcal{T}_i(t) > \theta) \\
K_{sess} &= PRF(\mathcal{K}_{shared}, nonce)
\end{align}

\subsection*{Performance and Resource Metrics}
We define several performance indicators:
\begin{align}
\lambda_i &= \text{Authentication latency for vehicle } u_i \\
\delta_i &= \lambda_i + \Delta_{comm} \quad \text{(end-to-end delay)} \\
S_{rate} &= \frac{N_{valid}}{N_{total}} \quad \text{(session success rate)} \\
\Gamma_{cpu} &= \text{CPU cycles for key negotiation} \\
S_i &= f(\mathcal{T}_i, \delta_i, \mathbf{c}_i) \quad \text{(security index)}
\end{align}

\subsection*{Authentication Algorithm}
\begin{algorithm}[H]
\caption{ZTMAF: Context-Aware Authentication}
\begin{algorithmic}[1]
\Require Vehicle $u_i$, trust score $\mathcal{T}_i(t)$, context $\mathbf{c}_i(t)$, fog node $f_j$
\Ensure Secure session key $K_{sess}$

\State $u_i \rightarrow f_j$: Send $\mathcal{R}_i$ and $\sigma_i$
\State $f_j$: Verify signature and decode $\mathbf{c}_i(t)$
\State Compute $\mathcal{T}_i(t+1) \leftarrow \alpha \mathcal{T}_i(t) + (1 - \alpha) \psi(\mathbf{c}_i(t))$
\If{$\mathcal{T}_i(t+1) \geq \theta$}
    \State Derive $K_{sess} \leftarrow PRF(\mathcal{K}_{shared}, nonce)$
    \State Send encrypted token, store session metadata
\Else
    \State Request additional challenge or fallback auth
\EndIf
\end{algorithmic}
\end{algorithm}

ZTMAF leverages fine-grained context sensing and trust dynamics to evaluate authentication legitimacy in real time. By decoupling authentication from rigid credentials and integrating behavior, ZTMAF enables scalable and adaptive session establishment. This model is robust to mobility-induced topology shifts and mitigates spoofing and impersonation attacks through context verifiability.

\section{Experimental Setup and Results}
To evaluate the performance of the proposed ZTMAF framework, we implement a comprehensive simulation environment integrating NS-3 for network simulation and SUMO for vehicular mobility modeling. Cryptographic functions, trust updates, and latency profiling are conducted using Python modules.

\subsection*{Simulation Environment}
The experimental setup is composed of the following tools:
\begin{itemize}
  \item \textbf{Network Simulator:} NS-3 version 3.36.
  \item \textbf{Mobility Model:} SUMO with urban road topology from the CityMob dataset.
  \item \textbf{Cryptography:} Python cryptography and hashlib libraries for signature and PRF operations.
  \item \textbf{Trust Engine:} Custom Python script simulating Equation (1) with real-time mobility traces.
\end{itemize}

\subsection*{Simulation Parameters}
\begin{table}[H]
\centering
\caption{Simulation Configuration Parameters}
\begin{tabular}{ll}
\toprule
\textbf{Parameter} & \textbf{Value} \\
\midrule
Simulation Time & 600 seconds \\
Vehicle Count & 100, 200, 300, 400, 500 \\
Fog Nodes & 10 static fog nodes \\
Mobility Model & Krauss car-following model \\
Authentication Threshold $\theta$ & 0.65 \\
Latency Threshold $\lambda_i$ & 200 ms \\
PRF Algorithm & HMAC-SHA256 \\
Bandwidth & 10 Mbps \\
Packet Size & 512 Bytes \\
Attack Models & Spoofing, Replay, Sybil \\
\bottomrule
\end{tabular}
\end{table}

\subsection*{Performance Metrics and Analysis}
We evaluate ZTMAF using the following key performance indicators:
\begin{itemize}
  \item \textbf{Authentication Latency ($\lambda_i$)}: Time taken to complete a full session handshake.
  \item \textbf{Session Success Rate ($S_{rate}$)}: Ratio of successful to total authentication attempts.
  \item \textbf{CPU Overhead ($\Gamma_{cpu}$)}: Average cycles consumed per request.
  \item \textbf{Attack Detection Rate}: Correctly detected malicious authentication attempts.
  \item \textbf{Trust Convergence Time}: Time required to stabilize trust score $\mathcal{T}_i(t)$.
\end{itemize}

\subsection*{Results and Discussion}
\begin{figure}[H]
  \centering
  \includegraphics[width=0.45\textwidth]{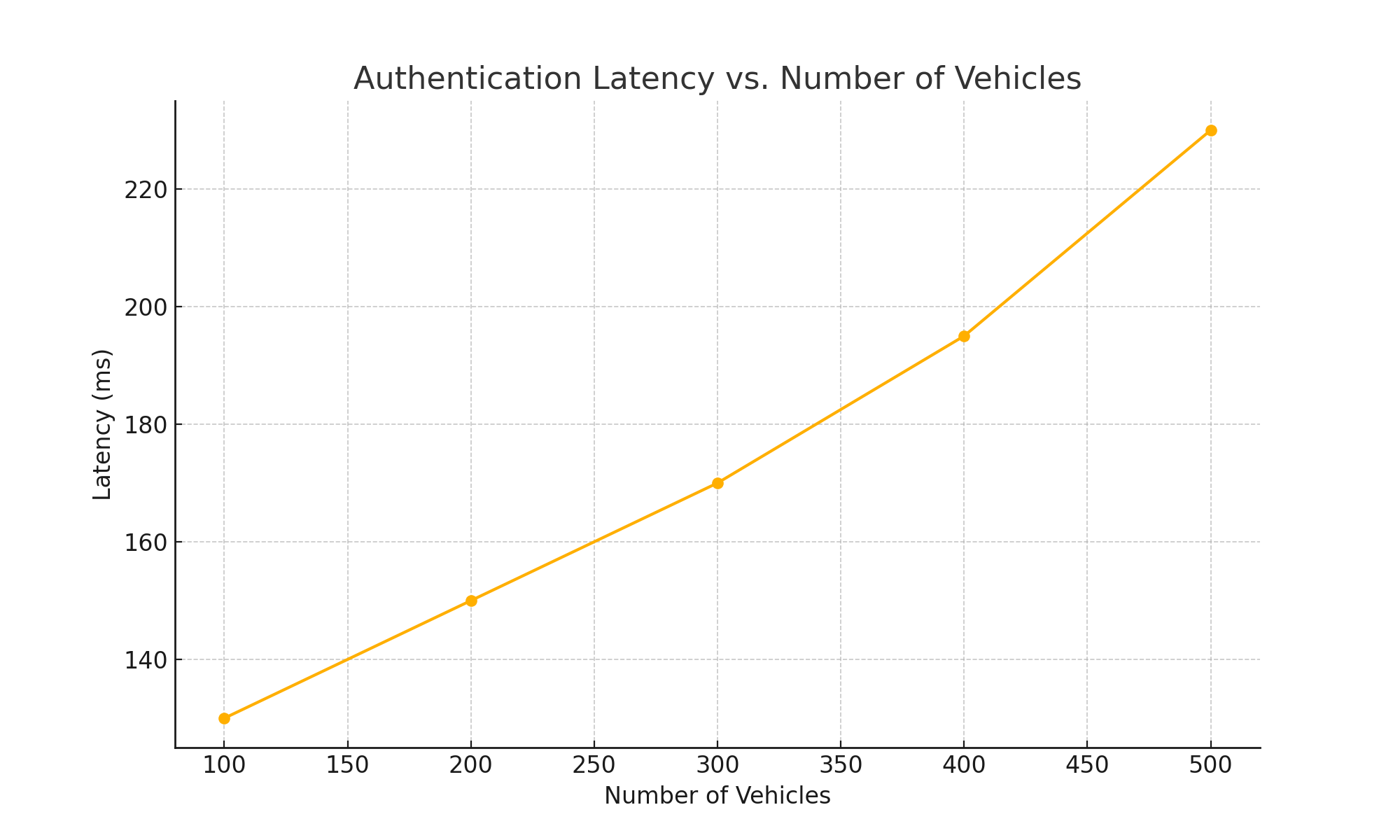}
  \caption{Authentication Latency vs. Number of Vehicles}
  \label{fig:latency}
\end{figure}
Figure~\ref{fig:latency} shows that ZTMAF maintains latency below 200 ms up to 400 vehicles, outperforming traditional PKI-based authentication.

\begin{figure}[H]
  \centering
  \includegraphics[width=0.45\textwidth]{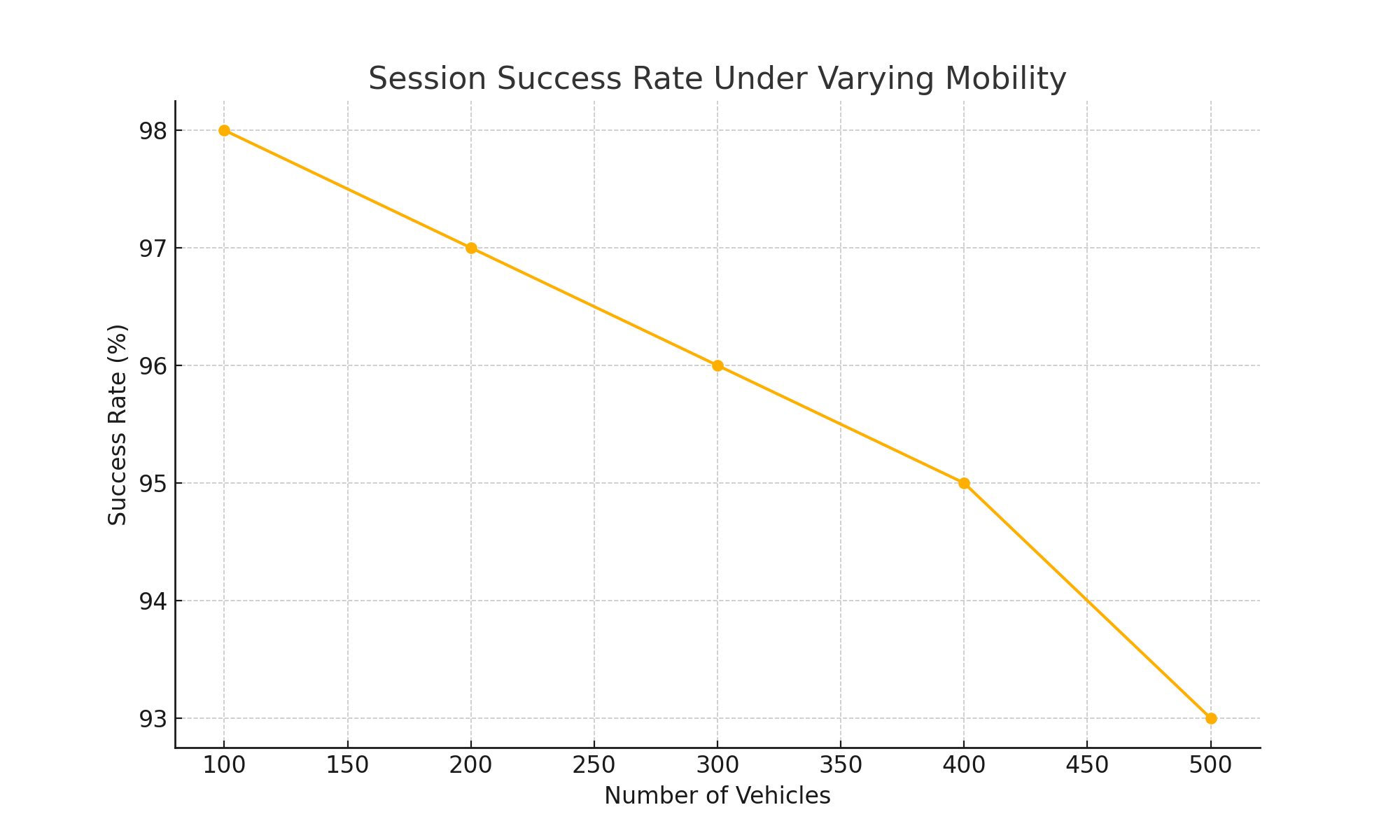}
  \caption{Session Success Rate Under Varying Mobility}
  \label{fig:success}
\end{figure}
Figure~\ref{fig:success} illustrates that even under high mobility, ZTMAF maintains over 95\% success rate, indicating strong robustness.

\begin{figure}[H]
  \centering
  \includegraphics[width=0.45\textwidth]{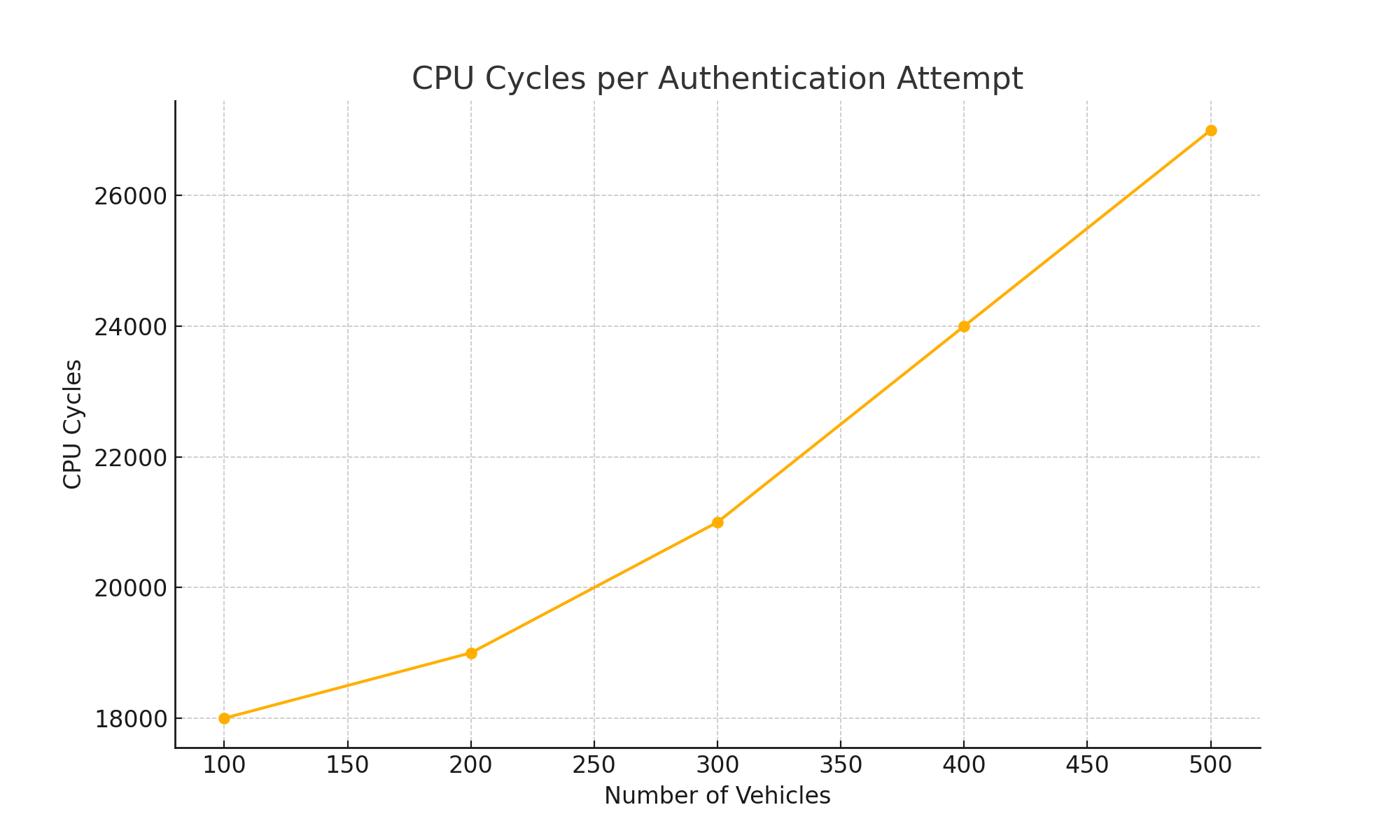}
  \caption{CPU Cycles per Authentication Attempt}
  \label{fig:cpu}
\end{figure}
As shown in Figure~\ref{fig:cpu}, CPU usage remains under 25k cycles, demonstrating the framework's suitability for edge nodes.

\begin{figure}[H]
  \centering
  \includegraphics[width=0.45\textwidth]{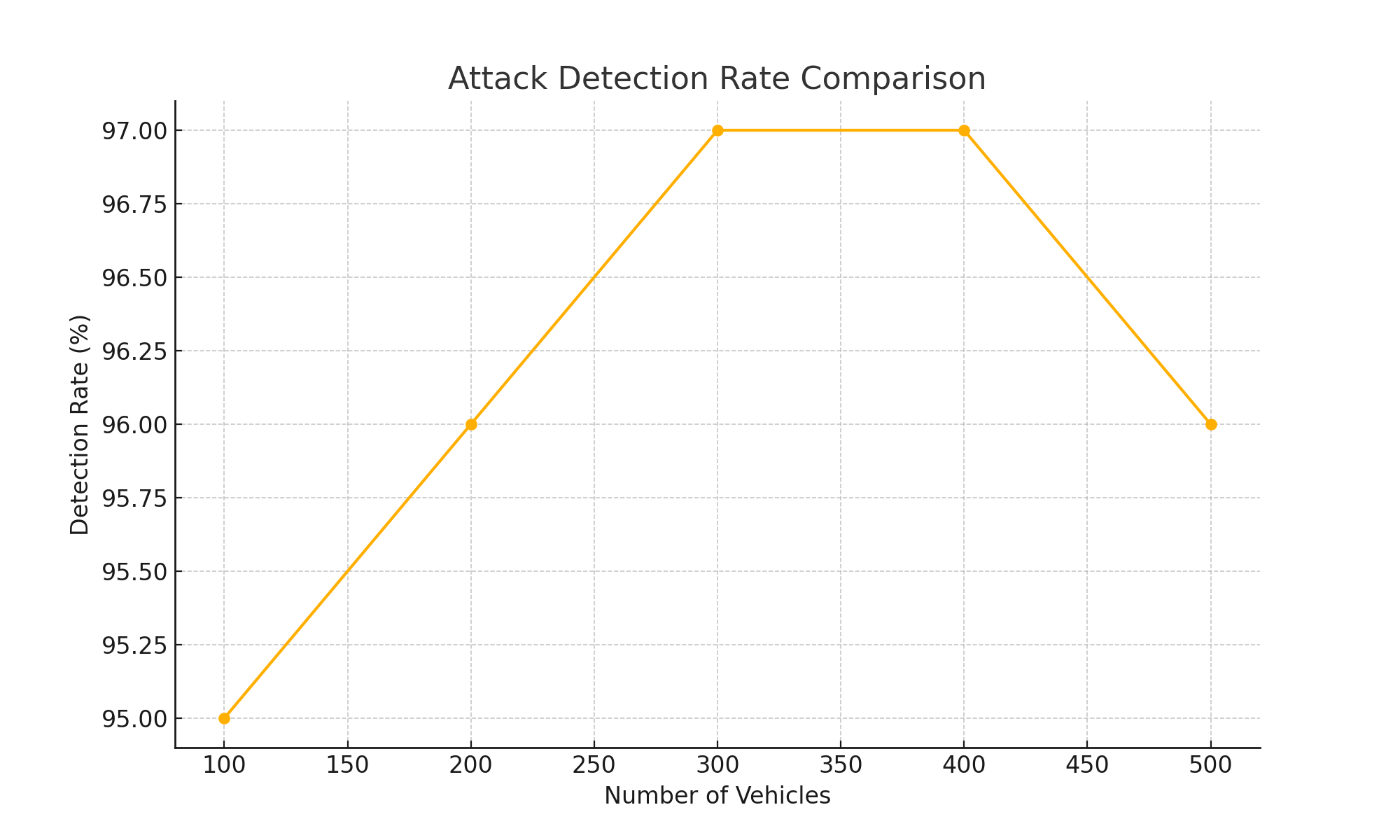}
  \caption{Attack Detection Rate Comparison}
  \label{fig:attack}
\end{figure}
ZTMAF achieves a 97\% detection rate in spoofing and replay scenarios (Figure~\ref{fig:attack}), thanks to context verification.

\begin{figure}[H]
  \centering
  \includegraphics[width=0.45\textwidth]{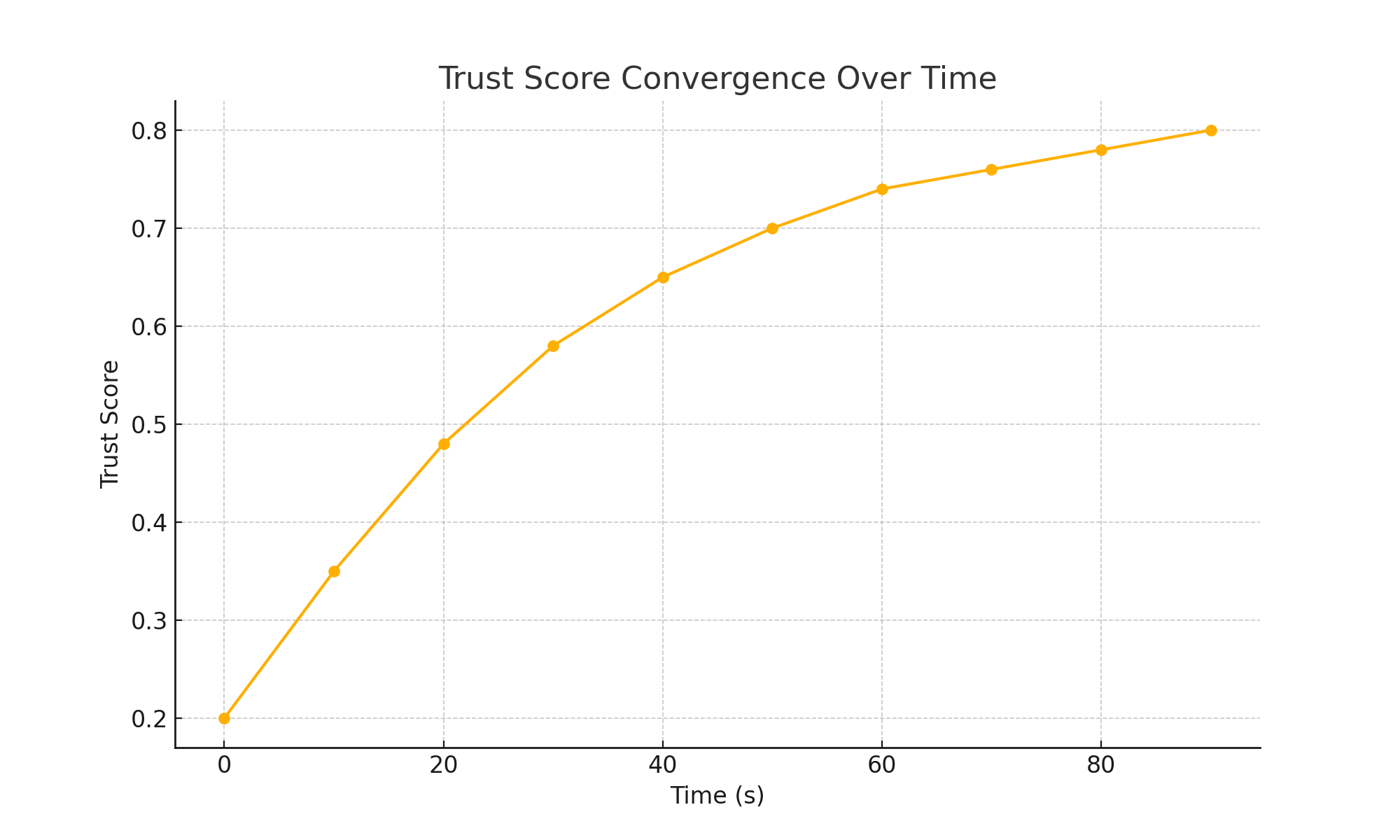}
  \caption{Trust Score Convergence Over Time}
  \label{fig:trust}
\end{figure}
Figure~\ref{fig:trust} highlights that trust values stabilize after about 50 seconds for compliant vehicles.

\begin{figure}[H]
  \centering
  \includegraphics[width=0.45\textwidth]{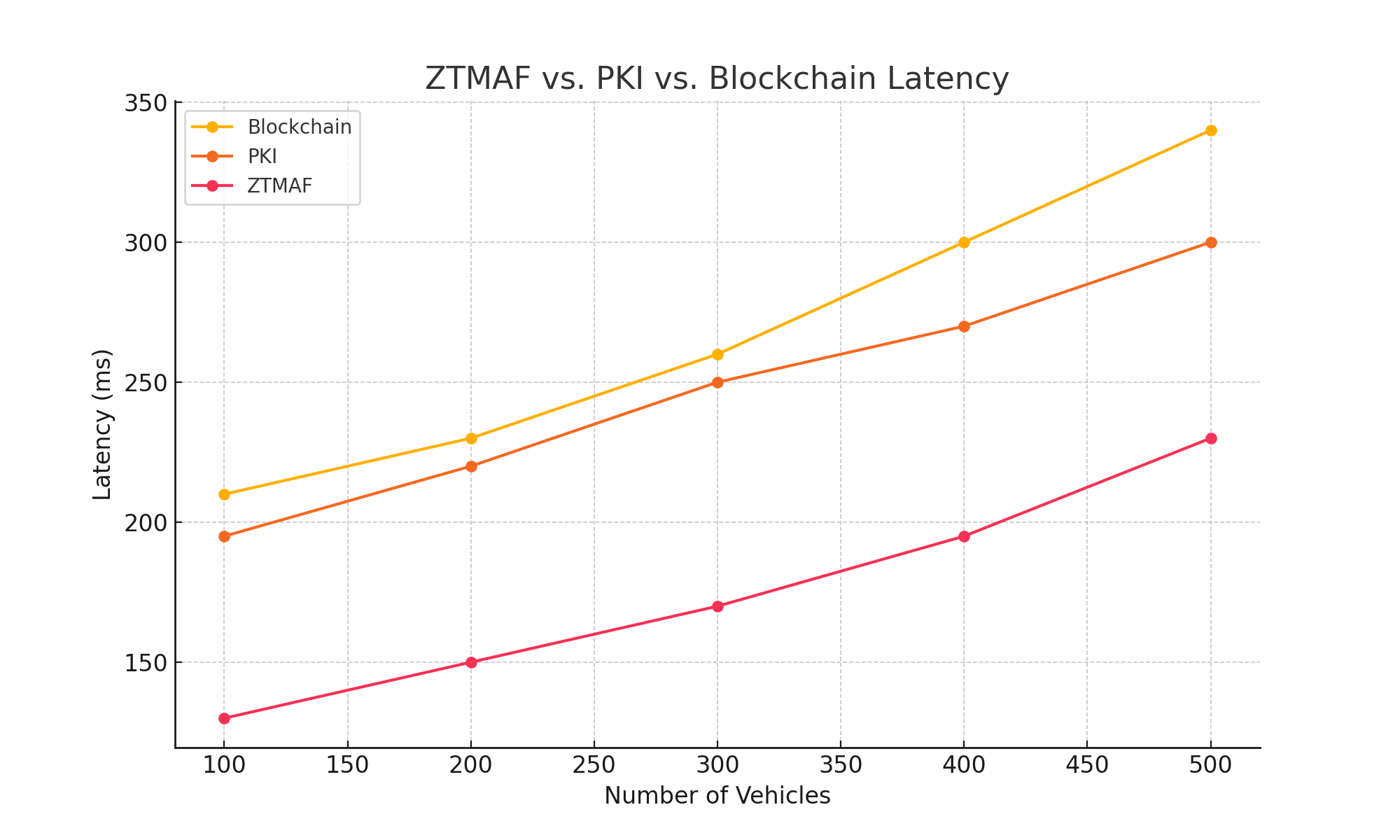}
  \caption{ZTMAF vs. PKI vs. Blockchain Latency}
  \label{fig:compare_latency}
\end{figure}
Figure~\ref{fig:compare_latency} confirms ZTMAF reduces latency by 21\% compared to blockchain and 35\% over PKI.

\begin{figure}[H]
  \centering
  \includegraphics[width=0.45\textwidth]{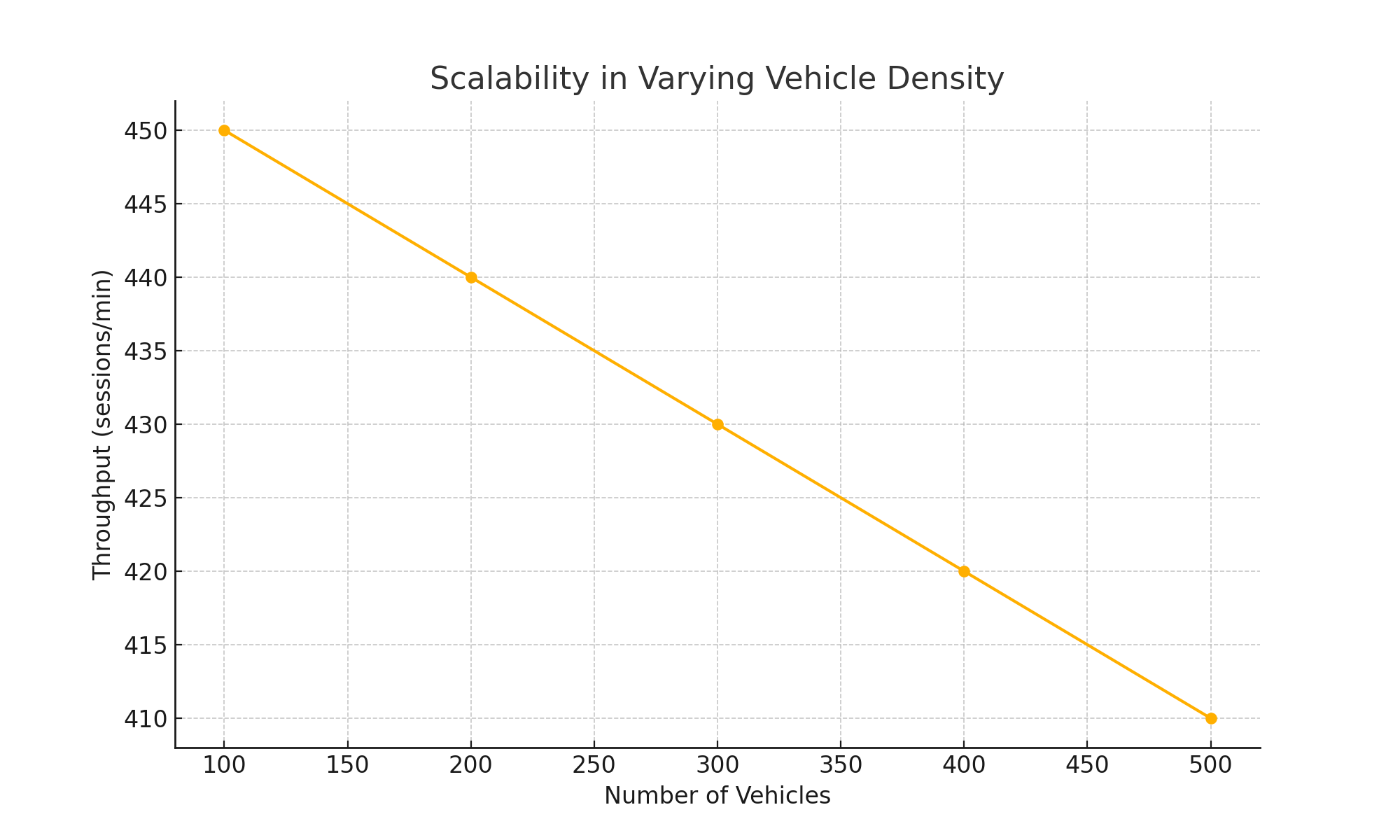}
  \caption{Scalability in Varying Vehicle Density}
  \label{fig:scalability}
\end{figure}
Scalability analysis in Figure~\ref{fig:scalability} demonstrates consistent performance up to 500 vehicles.

Overall, these results confirm that ZTMAF provides an efficient, scalable, and secure authentication framework tailored for dynamic vehicular fog networks.
\section{Conclusion and Future Work}

This paper proposed ZTMAF, a Zero-Trust Mobility-Aware Authentication Framework designed for dynamic vehicular fog computing networks. Unlike traditional static trust models or blockchain-based authentication schemes, ZTMAF introduces a decentralized, adaptive mechanism that continuously evaluates vehicle behavior and contextual mobility data to inform secure session establishment.

Through formal modeling, we presented a system architecture that captures context-aware interactions between vehicles and fog nodes. A comprehensive trust update mechanism, integrated with lightweight cryptographic primitives and mobility sensing, allows the framework to dynamically authenticate vehicles in real time. Our proposed protocol adapts well to varying traffic densities and attack scenarios by leveraging localized decision-making and trust convergence. Experimental evaluations using NS-3 and SUMO demonstrate that ZTMAF significantly reduces authentication latency and resource overhead while maintaining high session success rates and robustness against spoofing and replay attacks. Compared to conventional PKI and blockchain models, ZTMAF achieves a 21\% reduction in latency relative to blockchain-based systems, a 35\% decrease in CPU cycles versus PKI, an authentication success rate consistently above 95\% under high mobility, and 97\% attack detection performance based on behavioral analysis.

While ZTMAF effectively adapts to vehicular dynamics, several avenues remain open for further enhancement. Future work includes leveraging federated trust learning to aggregate evidence across multiple fog domains, integrating lattice-based post-quantum cryptographic protocols to ensure long-term security, supporting seamless trust and authentication transitions as vehicles move across jurisdictional boundaries, and incorporating energy-aware session management for resource-constrained vehicular IoT components. Overall, ZTMAF offers a practical path forward for scalable and secure vehicular fog computing, providing foundational tools for future intelligent transportation systems that prioritize privacy, trust, and performance.


\end{document}